# Numerical Simulation of the Mechanical Properties of Nanoscale Metal Clusters Using the Atomistic-Continuum Mechanics Method

Chan-Yen Chou, Cadmus Yuan, Chung-Jung Wu and Kuo -Ning Chiang*

Advanced Microsystem Packaging and Nano-Mechanics Research Lab
Dept. of Power Mechanical Engineering
National Tsing Hua University
101, Sec. 2, Kuang-Fu Rd.,
Hsinchu, Taiwan 300, R.O.C.

**Abstract**

A novel atomistic-continuum method (ACM) based on finite element method (FEM) is proposed to numerically simulate the nano-scaled Poisson's ratio and Young's modulus effect of Lithium (Li) body-centered cubic (BCC) structure. The potential energy between Li atoms is described by the Morse potential function [1]. The pre-force effect will be discussed due to the different Li lattice length between experimental lattice constant and diatom distance from Morse function. Moreover, the size effect of the nano-scaled Li cluster will be introduced.

## 1. Introduction

The unsaturated chemical bond of the surface molecule induces the physical and material properties at the solid surface to be different from those at the solid interior. In addition, because the ratio of the unsaturated surface atoms to the saturated atoms increases as the geometrical size decreases, the unsaturated surface atom effect is apparent at the nanoscale structure. Besides, the resolution of the measurement technology of nanoscale structures limits researchers in their precise understanding of the physical behavior of the nanoscale structure. For bulk-scaled metal, the mechanical properties can be represented by a simple material test (e.g. uniaxial stretching experiment), but the resolution of the atomistic experiment restricts researchers from thoroughly elucidating the mechanical response of a nanoscale metal. Accordingly, the quantitative mechanical properties of a nanoscale metal, such as Young's modulus and the Poisson's ratio, become difficult to achieve.

As a result, few feasible theoretical models have been developed successfully to describe the physical behavior in nanoscale material. Shenoy et al [2] proposed the mixed atomistic and continuum method to investigate the defect and the dislocation movement of the metal. Jeng et al [3] established a two-dimensional model of copper membrane, to simulate the deformation and the material properties of the nanoscale membrane under tensile loading. In addition, the Morse potential function is applied to describe the interatomic force. Yuan and Chiang [4] modeled the double-stranded deoxyribonucleic acid (dsDNA) molecule under axial and unzipping loading. Because the conventional molecule dynamic simulation technique induces an small increasing time-step which limits the total simulation time, the clustered atomistic-continuum method (CACM) based on the finite element method was applied to reduce the computational time.

In this research, the atomistic-continuum method (ACM) based on the finite element method was adopted to numerically simulate the mechanical response of the nanoscale Li metals. It is worth noting that the binding energy between Li atoms can be described by the Morse function, including the attractive and repulsive terms. The simulation results of the proposed model are accomplished by a commercial finite element solver. Due to the fact that the lattice length of Li at room temperature is not equal to the equilibrium length of two Li atoms, the pre-force effect is thoroughly discussed by numerical simulation and the bubble-like phenomenon is discovered. The size effect of the nanoscale Li cluster is considered as well. The simulation results indicate that the surface effect of the material significantly influences the material properties of the Li, and a 5nm$^3$ block will exhibit Young's modulus 10 times larger than the bulk one.

## 2. Fundamental Theories

### 2.1. Interatomic Potential Function

In this research, the Morse potential function [1] is used to describe the interatomic forces of Li atoms. The Morse potential function is an experiment based function, and it

---

*Corresponding Author, Professor, Dept. of Power Mechanical Engineering, National Tsing Hua University

describes the relationship between the bond energy and the bond length of the diatomic system.

The Morse function can be expressed as:

$$E(r_{ij}) = D\left(e^{-\alpha(r_{ij}-r_0)} - 1\right)^2 \tag{1}$$

where $E$ represents the potential energy of the diatomic system, $r_{ij}$ is the distance of atom $i$ and atom $j$, $D$ is the dissociation energy of the diatomic system, $r_0$ is the equilibrium length, and $\alpha$ is a characteristic constant which is determined by experimental results.

The Morse function describes the classical vibration of a chemical bond. The anharmonic Morse model provides a more realistic description of bond vibration than the harmonic potential functions. First, the Morse function describes the bond-breaking phenomenon of the chemical bond while it is over-stretched. In addition, the Morse function also precisely advises the dissociation energy of the potential energy. Furthermore, the mathematic characteristic of the Morse function describes the physical behavior when the ramping of the energy is more rapid under compression than it is under extension.

However, the parameters of the potential function are not always obtained for all kinds of atoms because of a lack of experimental results. Therefore, experimental theories have been established to develop specific potential functions which can, in principle, be extended to the entire periodic table for obtaining the parameters of those potential functions that have not been obtained experimentally. The Universal Force Field (UFF) theory [5] has been developed to estimate the parameters in the Morse function.

Based on the UFF theory, the Morse potential function parameters can be determined by force constant $k_{ij}$. A diatomic system can be considered as two atoms which are bonded by an effective spring with a force constant $k_{ij}$. The vibrational energy $E_{vib}$ of the system is then expressed as follows.

$$E_{vib} = \left(v + \frac{1}{2}\right)\hbar\omega \tag{2}$$

where $v = 1, 2, 3\ldots$, $\hbar = h/2\pi$. $h$ represents Planck's constant, and $\omega$ represents the oscillation frequency. The effective force constant $k_{ij}$, is determined from the vibrational energy.

$$k_{ij} = \mu\omega^2 = 664.12 \frac{Z_i^* Z_j^*}{r_{ij}^3} \tag{3}$$

where $\mu$ represents the effective mass of the diatomic system, $Z_i^*$ is the effective atomic charge of atom $i$, and $r_{ij}$ is the distance between atom $i$ and $j$. The parameter $\alpha$ in the Morse potential function is therefore determined by force constant $k_{ij}$.

$$\alpha = \left[k_{ij}/2D\right]^{\frac{1}{2}} \tag{4}$$

### 2.2. Atomistic-Continuum Mechanics Method

Based on the finite element method, the atomistic–continuum mechanics method is developed to simulate the mechanical characteristics, such as the Young's modulus and Poisson's ratio of nanoscale structures. The ACM method transfers an originally discrete atomic potential into an equilibrium continuum model by atomistic-continuum transfer elements. It simplifies the complexities of the interactive forces among the atoms, while keeping the calculation accuracy still acceptable and the computational time affordable.

One can generate the equations for a typical static constant-strain finite element. The total potential energy is a function of the nodal displacements $x$, $y$ and $z$ such that $\pi_p = \pi_p(x, y, z)$. Here the total potential energy is given by

$$\pi_p = U + \Omega_b + \Omega_p + \Omega_s \tag{5}$$

where $U$, $\Omega_b$, $\Omega_p$ and $\Omega_s$ represent the strain energy, the potential energy of the body force, the potential energy of the concentrated load and the potential energy of the distributed load, respectively. The above equation can be rewritten as a finite element integrated form:

$$\begin{aligned}\pi_p = &\frac{1}{2}\iiint_V \{d\}^T[B]^T[D][B]\{d\}dV - \\ &\iiint_V \{d\}^T[N]^T\{F\}dV - \{d\}^T\{P\} - \\ &\iint_S \{d\}^T[N_s]^T\{T_s\}dS\end{aligned} \tag{6}$$

where $\{d\}$ represents the nodal displacement vector, $[B]$ is the strain-displacement matrix, $[D]$ is the modulus of the elasticity matrix, $[N]$ is the shape function matrix, $\{F\}$ is the body force vector, $\{P\}$ is the external load vector and $\{T_s\}$ is the traction force vector.

The ACM method transfers the interatomic potential function into a force-displacement curve so as to create an equivalent atomistic-continuum transfer element. Afterwards, the equivalent nanoscale model can be analyzed by FEM.

### 3. Modeling of Metal Crystal via ACM

Based on the ACM theories described above, one can analyze the mechanical behavior of the crystal structure. In this research, one of the alkali metals, the Lithium BCC

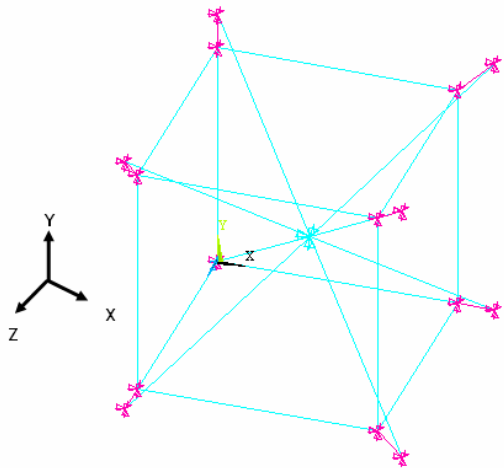

Fig. 1 The ACM model of the single BCC crystal preforce analysis

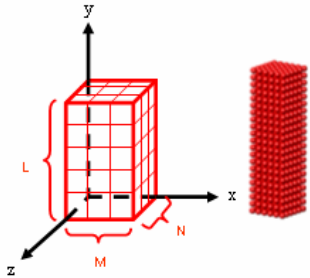

Fig. 2 The ACM model of a $N \times M \times L$ cubic structure.

structure is used as the test vehicle to verify the feasibility of the ACM theory.

The Morse potential function is utilized herein to describe the interatomic forces of Li metal. The lattice constant of Li is 3.51Å [5] and the dissociation energy of Li is 1.66eV [6]. In addition, the basic atomic structure of Li metal is a body centered cubic (BCC), which consists of one body atom and eight corner atoms in one cubic.

In this research, the interatomic forces between body center atom and corner atom (inner elements) and those between-corner-atoms (outer elements) are considered as ACM transfer elements. Therefore, there are 8 inner elements and 12 outer elements in one BCC structure. Moreover, a nano-scaled $N \times M \times L$ BCC cubic structure could be constructed based on the single BCC structure as shown in Fig. 2.

### 4. Results and Discussion

The finite element method based commercial program ANSYS® is utilized to investigate the preforce performance of Li BCC structure. Two models of inner elements and outer elements are constructed separately, and all elements are of length $2R$. Then, all related nodes in the inner elements model and outer element model are tied together using appropriate constraint equations. Fig. 1 represents the ACM model in which the blue lines represent the ACM elements, the purple arrows represent the constraint equations and the blue arrows represent the boundary conditions defining the displacement of the body center atom equaling to zero. In the single BCC crystal analysis, the preforce of inner elements was 648 pN in compression, and 374 pN in tension for the outer elements.

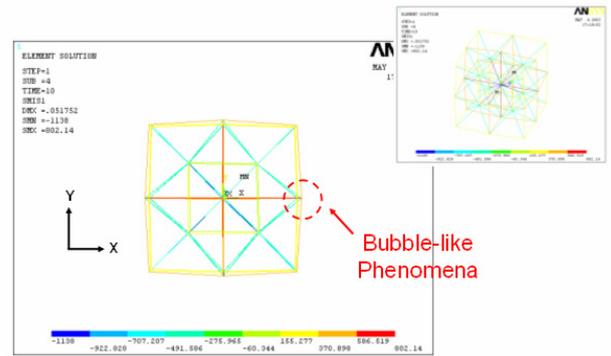

Fig. 3 The bubble-like phenomena of the preforce analysis.

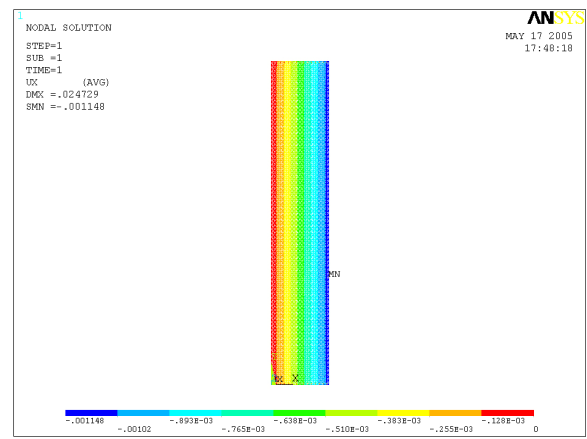

Fig. 4 The ACM model of an $n \times m \times l$ cubic structure and its results.

In addition, a bubble-like phenomenon was observed when the ACM model extended to a $2 \times 2 \times 2$ BCC cubic structure, as shown in Fig. 3. The bubble-like phenomenon is induced by the surface atom with out-of-plane displacement. Moreover, the preforce distribution is non-uniform in the $2 \times 2 \times 2$ cubic ACM model, indicating that the assumption of uniformly distributed stretching in the analytical solution may not be suitable for the preforce analysis.

The ACM model of an $n \times m \times l$ BCC cubic

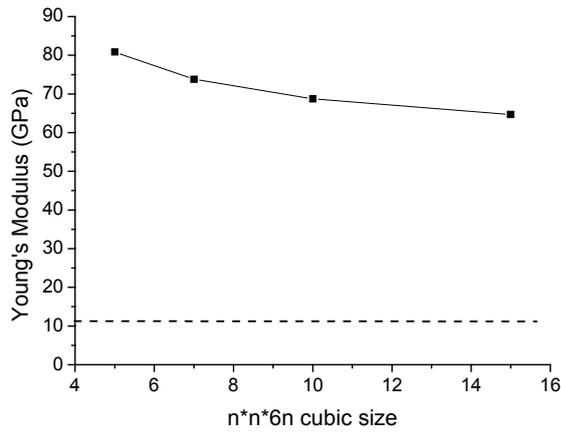

Fig. 5 Young's modulus of nanoscale structures with constant aspect ratio

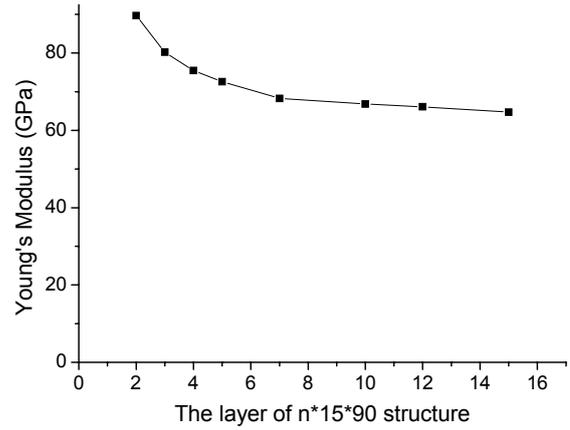

Fig. 7 Surface effect analysis of the Young's modulus of nanoscale structures.

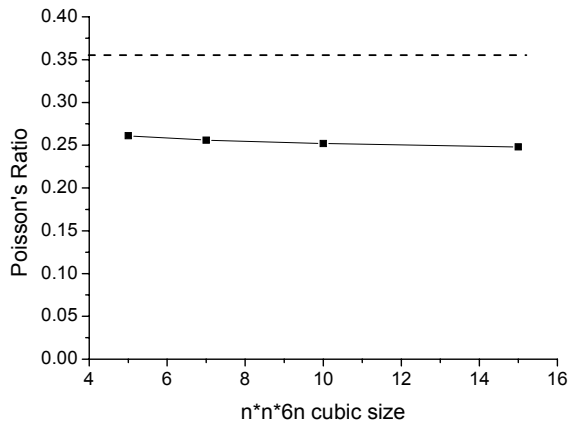

Fig. 6 Poisson's ratio of nanoscale structures with constant aspect ratio

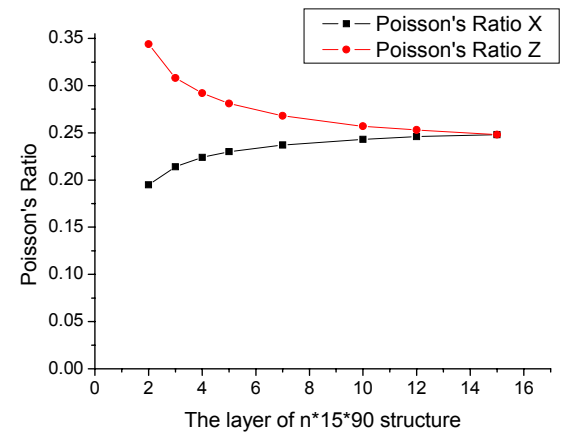

Fig. 8 Surface effect analysis of Poisson's ratio of nanoscale structures.

structure and its simulation results are shown in Fig. 4. The ACM model is applied by a prescribed displacement loading in the Y direction. Consequently, the nanoscale BCC structure necks down in the middle section and the reaction forces in the Y direction are obtained. In this research, the strain of the nanoscale structure $\varepsilon_x$, $\varepsilon_y$, $\varepsilon_z$ is determined by the maximum displacement dividing the original length at the $X$, $Y$, $Z$ direction individually. Furthermore, the Poisson's ratio of the nanoscale structure $v_x$, $v_z$ is defined as $v_x = -(\varepsilon_x/\varepsilon_y)$, $v_z = -(\varepsilon_z/\varepsilon_y)$. In addition, the Young's modulus is defined as $E = \sigma_y/\varepsilon_y$ in which the Y direction stress $\sigma_y$ is determined by the total reaction force dividing the cross-section of the ACM model. Based on the foregoing definitions, the Young's modulus and the Poisson's ratio of a nanoscale structure can therefore be investigated. The following topics investigate the size-dependent material properties of a nanlscale structure.

To investigate the mechanical properties of a nanoscale structure, four ACM models with constant aspect ratio of *5 × 5 × 30, 7 × 7 × 42, 10 × 10 × 60,* and *15 × 15 × 90* BCC cubic structure are constructed to evaluate Young's modulus and Poisson's ratio of these models. The ratio of the surface atoms to the inner atoms are 53.7%, 36.1%, 24.1% and 15.5%, respectively.

The simulation results are shown in Fig. 5 and Fig. 6, with the dashed line indicating the bulk properties of the Li metal [7]. The Young's modulus of all nanoscale structures is about seven times larger than that of the bulk property. When structures shrink to nanoscale, the ratio of unsaturated surface atoms to saturated inner atoms increases rapidly. Therefore, the surface atoms strengthen the nanoscale structure due to the cohesion of the surface atoms. Since no random material defects or dislocations are considered in this research, the ACM model results in a higher Young's modulus than the bulk Li metal.

In addition, the Young's modulus drops as the size of the nanoscale structure increases, because the ratio of surface atoms to inner atoms decreases. Therefore, the material properties of the nanoscale structure are highly dependent upon the size of the nano cluster.

Accordingly, the cohesion of the surface atom decreases and thereby weakens the nanoscale structure. The bulk material properties apparently do not fluctuate by the size effect because the ratio of surface atoms to inner atoms equals nearly zero. As a result the inner atoms generally dominate the material properties of the bulk structure.

The surface effect is further discussed from the thin film lattice to the cubic structure. We first assume a fixed area of 15 × 90 lattice area size in the XY plane, in which an ACM model of *2 × 15 × 90* structure is constructed. Seven models based on the previous one, up to a *15 × 15 × 90* BCC cubic structure are then analyzed, and the Young's modulus and Poisson's ratio of these models are evaluated.

The simulation results are shown in Fig. 7 and Fig. 8. The Young's modulus drops when the width of the nanoscale structure increases. However, the Young's modulus drops slowly after the *7 × 15 × 90* cubic model which indicates that the surface atom effect in the Z direction become smaller. Another interesting phenomenon observed in this analysis is that the Poisson's ratio becomes anisotropic. The increase of the width of the nanoscale structure decreases the X direction's Poisson's ratio $v_x$. In other words, the surface atom effect in the Z direction vanishes after the *7 × 15 × 90* cubic model.

## 5. Conclusion

This paper proposed a novel atomistic- continuum method based on the finite element method to numerically simulate the material properties of the nanoscale structure. The Li BCC structure was taken as a test vehicle to validate the flexibility of the proposed ACM method.

Because of the numerical characteristics of the ACM method, the computing effort to accomplish a specific ACM model simulation is much less than the conventional molecular dynamics method. However, the degree of accuracy of the ACM simulation results is the same as the MD ones.

Moreover, the preforce analysis indicates that there are compressive preforces between the body center atoms and the corner atoms in the stable Li BCC structure, with tensile preforces between the corner atoms. In addition, a bubble-like phenomenon can be observed by static BCC structural analysis, indicating an out-of-plane displacement of the surface atoms.

The mechanical properties analysis of the nanoscale structure indicates that the Young's modulus of nanoscale structures is much larger than that of the bulk property because the material defects are not considered. Furthermore, the simulation results indicate that the material properties of nanoscale structures are highly size-dependent, and this phenomenon is caused by the different ratio of surface atoms to inner atoms.